\documentclass[aps,pra,twocolumn,showpacs,floatfix]{revtex4}
\usepackage{graphicx}
\usepackage{nicefrac}
\usepackage{amsmath}
\usepackage{amsfonts}
\usepackage{amssymb}
\usepackage{amsthm}
\usepackage{epsf}
\usepackage{bm}
\usepackage{bbm}
\usepackage{longtable}

\usepackage{dcolumn}

\newcolumntype{.}{D{x}{}{-1}}

\newcommand{\bsigma}{\vec{\sigma}}

\newcommand{\bnabla}{\vec{\nabla}}

\newcommand{\bfr}{\vec{r}}

\newcommand{\bfp}{\vec{p}}

\newcommand{\Za}{{Z\alpha}}

\newcommand{\cE}{{\cal E}}

\newcommand{\lbr}{\left<} \newcommand{\rbr}{\right>}

\begin{document}

\title{Fine structure of helium-like ions and determination of the 
fine structure constant}

\author{Krzysztof Pachucki} \affiliation{Institute of Theoretical
        Physics, University of Warsaw, Ho\.{z}a 69, 00--681 Warsaw, Poland}

\author{Vladimir A. Yerokhin} \affiliation{Center for Advanced Studies,
        St.~Petersburg State Polytechnical University, Polytekhnicheskaya 29,
        St.~Petersburg 195251, Russia}

\begin{abstract}
We report a calculation of the fine structure splitting in light
helium-like atoms, which accounts for all quantum electrodynamical effects up to
order $\alpha^5\,$Ry. For the helium atom, we resolve the previously reported disagreement
between theory and experiment and determine the
fine structure constant with an accuracy of 31~ppb. The calculational results
are extensively checked by comparison with the experimental data for different
nuclear charges and by evaluation of the hydrogenic limit of individual corrections.
\end{abstract}


\maketitle

Accurate measurements of the fine structure of the $2^3P$ level of helium
and helium-like ions make possible a precise test of quantum electrodynamic
(QED) theory of the electron-electron interaction in bound systems. 
Alternatively, assuming the validity of the theory, the
fine structure constant $\alpha$ can be determined with a high accuracy. This fact was
first pointed out by Schwartz in 1964 \cite{schwartz:64}. 
Fourteen years later, after a series of dedicated studies, 
Schwartz's program of calculations resulted in a
theoretical description of the helium fine structure complete up to order
$m\alpha^6$ (or $\alpha^4$~Ry) and a value of $\alpha$ accurate to
0.9~ppm \cite{lewis:78}. 

Further theoretical progress met serious difficulties. It was only in 1996
that a calculation of the dominant part of the next-order 
$m\alpha^7$ contribution was reported \cite{zhang:96:prl}.
To complete the calculation of this contribution turned out 
to be a challenge. A number of investigations 
\cite{pachucki:00:jpb,drake:02:cjp,pachucki:03:jpb} reported 
partial results, yielding significant disagreement with the
experimental data. The first complete calculation \cite{pachucki:06:prl:he} 
increased the disagreement even further by reporting differences of 
more than 10 standard deviations 
with the experimental results for the $2^3P_0-2^3P_1$ ($=\nu_{01}$) 
and $2^3P_1-2^3P_2$ ($=\nu_{12}$) intervals of helium 
\footnote{
In the present Letter, this disagreement is explained to be due to a sign
mistake in the original calculation of Bethe logarithms \cite{pachucki:00:jpb} 
and an inaccuracy in the nuclear recoil contribution.}.

In our previous investigation \cite{pachucki:09:hefs} we 
recalculated all effects up to order $m\alpha^7$  to the fine structure of helium
with improved numerical precision,
and significantly reduced the deviation of theory from experiment. In this
Letter we eliminate a small inconsistency in our previous evaluation of Bethe
logarithms and obtain agreement with the latest experimental
results for helium. We also calculate the fine structure of helium-like 
ions with nuclear charges $Z$ up to $10$ and observe good agreement with most
of the experimental data. As an independent check of our calculations, we study 
the hydrogenic ($Z\to\infty$) limit of individual corrections 
and demonstrate the consistency of the obtained results with the hydrogen theory. 

The agreement observed for helium-like ions and the confirmed hydrogenic limit
are substantial evidences of the reliability of our helium results. 
We are thus in a position to make an independent
determination of the fine structure constant. The comparison of our
theoretical prediction for the $\nu_{01}$ interval in helium (accurate to
57~ppb) with the experimental result \cite{zelevinsky:05} (accurate to 
24~ppb) determines the value of $\alpha$ with an accuracy 
of 31~ppb, see Eq. (\ref{alpha}) below. 
This is currently the third-precise method of determination of $\alpha$, 
after the electron $g$ factor \cite{hanneke:08} and the atomic recoil effect
\cite{cadoret:08}. Measurements of $\alpha$ by different methods provide
a sensitive test of consistency of theory across a range of energy scales and
physical phenomena. 

\begin{table*}
\caption{Contributions of order $m\alpha^7$ and $m^2\alpha^6/M$
to the $2^3P_J-2^3P_{J'}$ fine-structure intervals of helium-like atoms. 
\label{tab:E7}} 
\begin{ruledtabular}
  \begin{tabular}{cl.....}
 $(J,J')$ &  
 $Z$ &  
\multicolumn{1}{c}{$\cE^{(6)}_{M}/[Z^8\,m/M]$}        
             &  \multicolumn{1}{c}{$\cE^{(7)}_{\rm log}/[Z^6\,\ln (\Za)^{-2}]$}        
                         &\multicolumn{1}{c}{$\cE^{(7)}_{\rm first}/Z^6$}        
                                  &\multicolumn{1}{c}{$\cE^{(7)}_{\rm sec}/Z^7$}        
                                         &\multicolumn{1}{c}{$\cE^{(7)}_{L}/Z^6$}        
    \\
    \hline\\[-5pt]
(0,1)
 &  2 &   -0.015x\,21  &    0.001x\,105\,3 &    0.002x\,213\,4 &    0.001x\,169\,3 &   -0.002x\,388\,1(1) \\
 &  3 &   -0.020x\,60  &   -0.001x\,149\,0 &    0.004x\,426\,9 &    0.001x\,581\,8 &    0.005x\,524\,0(1) \\
 &  4 &   -0.023x\,06  &   -0.001x\,846\,4 &    0.005x\,403\,0 &    0.001x\,906\,7 &    0.008x\,307\,0(1) \\
 &  5 &   -0.024x\,39  &   -0.001x\,836\,2 &    0.005x\,842\,6 &    0.002x\,158\,9 &    0.008x\,709\,1(1) \\
 &  6 &   -0.025x\,22  &   -0.001x\,593\,2 &    0.006x\,047\,0 &    0.002x\,357\,8 &    0.008x\,270\,6(1) \\
 &  7 &   -0.025x\,81  &   -0.001x\,287\,7 &    0.006x\,139\,9 &    0.002x\,518\,6 &    0.007x\,560\,6(1) \\
 &  8 &   -0.026x\,24  &   -0.000x\,980\,6 &    0.006x\,176\,7 &    0.002x\,651\,4 &    0.006x\,793\,0(1) \\
 &  9 &   -0.026x\,58  &   -0.000x\,693\,1 &    0.006x\,184\,0 &    0.002x\,763\,1 &    0.006x\,049\,1(1) \\
 & 10 &   -0.026x\,84  &   -0.000x\,431\,5 &    0.006x\,175\,6 &    0.002x\,858\,2 &    0.005x\,357\,9(1) \\
 &$\infty$[extrap.]  & 
          -0.029x\,4   &    0.003x\,315    &    0.005x\,415\,7 &    0.004x\,045\,2 &   -0.005x\,095\\
 &$\infty$[exact]    & 
                     &      0.003x\,316    &    0.005x\,415\,7 &    0.004x\,045\,2 &   -0.005x\,099
\\[5pt]
(0,2)
 &  2 &   -0.001x\,235 &    0.001x\,025\,6 &    0.003x\,016\,7 &   -0.000x\,393\,6 &   -0.001x\,716\,1(1) \\
 &  3 &   -0.000x\,418 &   -0.002x\,365\,8 &    0.007x\,084\,4 &   -0.001x\,857\,6 &    0.010x\,589\,2(1) \\
 &  4 &   -0.000x\,200 &   -0.002x\,947\,8 &    0.009x\,544\,9 &   -0.002x\,219\,8 &    0.014x\,039\,4(1) \\
 &  5 &   -0.000x\,069 &   -0.002x\,416\,4 &    0.011x\,062\,7 &   -0.002x\,222\,6 &    0.013x\,743\,0(1) \\
 &  6 &    0.000x\,006 &   -0.001x\,587\,4 &    0.012x\,062\,8 &   -0.002x\,119\,2 &    0.012x\,256\,1(1) \\
 &  7 &    0.000x\,045 &   -0.000x\,731\,5 &    0.012x\,760\,9 &   -0.001x\,988\,8 &    0.010x\,475\,7(1) \\
 &  8 &    0.000x\,066 &    0.000x\,066\,1 &    0.013x\,271\,0 &   -0.001x\,858\,0 &    0.008x\,716\,4(1) \\
 &  9 &    0.000x\,072 &    0.000x\,783\,4 &    0.013x\,657\,8 &   -0.001x\,735\,7 &    0.007x\,083\,6(1) \\
 & 10 &    0.000x\,074 &    0.001x\,420\,7 &    0.013x\,959\,9 &   -0.001x\,624\,3 &    0.005x\,604\,8(1) \\
 &$\infty$[extrap.]    & 
          -0.000x\,03  &    0.009x\,945    &    0.016x\,2473   &    0.000x\,000\,8 &   -0.015x\,283\\
 &$\infty$[exact]      & 
               0x      &    0.009x\,947    &    0.016x\,2471   &        0x         &   -0.015x\,296
  \end{tabular}
\end{ruledtabular}
\end{table*}

The energy levels of light atoms are addressed here within 
a rigorous QED approach based on
an expansion of both relativistic and radiative effects
in powers of $\alpha$ \cite{lepton:09}. 
This approach allows one to consistently 
improve the accuracy of calculations by accounting for various
effects order by order. The helium fine-structure splitting is thus 
represented as 
\begin{equation} \label{e1} 
E = m
\left[\alpha^4\cE^{(4)} + \alpha^5\cE^{(5)} + \alpha^6\cE^{(6)}
  + \alpha^7\cE^{(7)} + \ldots\right],
\end{equation}
where the expansion terms $\cE^{(n)}$ may include $\ln\alpha$. 
The summary of results for energy levels up to 
order of $m\alpha^6$ is given in
our previous investigation \cite{yerokhin:10:helike}. 
In the present Letter we evaluate corrections of order
$m\alpha^7$ and $m^2\alpha^6/M$, where $M$ is the nuclear mass.
The $m\alpha^7$ correction can be represented as a sum of four parts,
\begin{align} \label{ma7}
\cE^{(7)} = \cE^{(7)}_{\rm log}+ \cE^{(7)}_{\rm first} + \cE^{(7)}_{\rm sec}+ \cE^{(7)}_{L}\,.
\end{align} 
The first part combines all terms with $\ln Z$ and $\ln \alpha$
\cite{zhang:96,zhang:96:prl,pachucki:99:jpb}, 
\begin{align} \label{E7log}
\cE^{(7)}_{\rm log} &\ =  
\ln[(Z\,\alpha)^{-2}] \,\left[
\lbr \frac{2 Z}{3}\,
i\,\bfp_1\times\delta^3(r_1)\,\bfp_1\cdot\bsigma_1 \rbr
  \right.
   \nonumber \\ &
-\lbr \frac{1}{4}
   (\bsigma_1\cdot\bnabla)\,(\bsigma_2\cdot\bnabla) \delta^3(r)\rbr
-\lbr \frac{3}{2}\,
i\,\bfp_1\times\delta^3(r)\,\bfp_1\cdot\bsigma_1\rbr 
   \nonumber \\ &
   \left.
 + \frac{8Z}{3} \lbr H^{(4)}_{\rm fs} \frac1{(E_0-H_0)'} \bigl[
 \delta^3(r_1)+\delta^3(r_2)\bigr]\rbr 
\right] \,,
\end{align}
where $\vec{r} = \vec{r}_1-\vec{r}_2$, 
$H_0$ and $E_0$ are the Schr\"odinger Hamiltonian and its eigenvalue,
and $H^{(4)}_{\rm fs}$ is the spin-dependent part of the Breit-Pauli
Hamiltonian (see Eq.~(3) of Ref.~\cite{pachucki:09:hefs}). 

The second part of $\cE^{(7)}$ 
is induced by effective Hamiltonians to order $m\alpha^7$, 
which were derived by one of us 
(K.P.) in Refs. \cite{pachucki:06:prl:he,pachucki:09:hefs}. 
(The previous derivation of this correction
by Zhang \cite{zhang:96} turned out to be not entirely consistent.)
The result is
\begin{align}
\cE^{(7)}_{\rm first} = \Bigl<  H_Q + H_H + H^{(7)}_{\rm fs, amm} \Bigr>\,,
\end{align}
where the Hamiltonian $H_Q$ is induced by the two-photon exchange
between the electrons, the electron self-energy, and the vacuum polarization, 
$H_H$ represents the anomalous magnetic moment (amm) correction
to the Douglas-Kroll operators
(see Eq.~(101) of Ref.~\cite{pachucki:09:hefs}), and 
$H^{(7)}_{\rm fs, amm}$ is the $m\alpha^7$
part of the Breit-Pauli Hamiltonian with inclusion of the amm effect
(see Eq.~(3) of Ref.~\cite{pachucki:09:hefs}). The Hamiltonian $H_Q$
is 
\begin{align} \label{HQ}
H_Q  &\ = Z\,\frac{91}{180} 
\,i\,\bfp_1\times\delta^3(r_1)\,\bfp_1\cdot\bsigma_1
  \nonumber \\ &
-\frac12\, (\bsigma_1\cdot\bnabla)\,(\bsigma_2\cdot\bnabla)\, \delta^3(r)\,
 \left[ \frac{83}{30}+\ln Z\right]
 \nonumber \\ &
+3\,i\,\bfp_1\times\delta^3(r)\,\bfp_1\cdot\bsigma_1\,
  \left[ \frac{23}{10}-\ln Z\right]
 \nonumber \\ &
-\frac{15}{8\,\pi}\, \frac{1}{r^7}\,(\bsigma_1\cdot\bfr)\,(\bsigma_2\cdot\bfr)
-\frac{3}{4\,\pi}\,i\,\bfp_1\times \frac{1}{r^3}\, \bfp_1\cdot\bsigma_1\,.
\end{align}
Here, the terms with $\ln Z$ compensate 
the logarithmic dependence implicitly present in 
expectation values of singular operators $1/r^3$ and $1/r^5$.

\begin{table*}
\caption{Individual contributions 
to the $2^3P_J-2^3P_{J'}$ fine-structure intervals of helium-like atoms, in MHz$/Z^4$.
\label{tab:total}} 
\begin{ruledtabular}
  \begin{tabular}{cl......}
 $(J,J')$ &  
 $Z$ &  
\multicolumn{1}{c}{$m\alpha^4$}        
             &  \multicolumn{1}{c}{$m\alpha^5$}        
                         &\multicolumn{1}{c}{$m\alpha^6$}        
                                  &\multicolumn{1}{c}{$m\alpha^7({\rm log})$}        
                                         &\multicolumn{1}{c}{$m\alpha^7({\rm nlog})$}        
                                               &\multicolumn{1}{c}{Total}    \\
    \hline\\[-5pt]
(0,1)
 &  2 &  1847.x73534 &   3.41x900 &  -0.1x0109 &   0.00x509 &   0.00x118 & 1851.x05952(11) \\
 &  3 &  1917.x79396 &   3.24x978 &   1.2x278  &  -0.01x076 &   0.01x801 & 1922.x27881(59) \\
 &  4 &  1346.x96534 &   1.94x384 &   4.5x603  &  -0.02x843 &   0.04x648 & 1353.x4875(39) \\
 &  5 &   765.x88557 &   0.68x551 &  10.3x60   &  -0.04x139 &   0.08x628 &  776.x976(14) \\
 &  6 &   270.x38772 &  -0.36x757 &  19.2x39   &  -0.04x886 &   0.13x952 &  289.x349(37) \\
 &  7 &  -139.x08557 &  -1.22x955 &  31.8x63   &  -0.05x110 &   0.20x903 & -108.x294(83) \\
 &  8 &  -477.x53446 &  -1.93x791 &  48.9x20   &  -0.04x855 &   0.29x785 & -430.x30(17) \\
 &  9 &  -759.x77039 &  -2.52x632 &  71.1x10   &  -0.04x163 &   0.40x916 & -690.x82(31) \\
 & 10 &  -997.x72326 &  -3.02x103 &  99.1x20   &  -0.03x076 &   0.54x619 & -901.x11(53) 
\\[5pt]
(0,2)
 &  2 &  1992.x75043 &   2.00x994 &  -0.5x0717 &   0.00x472 &   0.00x028 &  1994.x25820(11) \\
 &  3 &  1150.x27490 &  -0.94x285 &  -0.8x6465 &  -0.02x216 &   0.01x483 &  1148.x46007(41) \\
 &  4 &  -384.x65915 &  -4.44x824 &  -1.3x897  &  -0.04x539 &   0.03x204 &  -390.x5104(12) \\
 &  5 & -1739.x32853 &  -7.32x066 &  -2.3x939  &  -0.05x446 &   0.04x661 & -1749.x0509(32) \\
 &  6 & -2838.x55028 &  -9.58x033 &  -3.9x945  &  -0.04x868 &   0.05x688 & -2852.x1169(77) \\
 &  7 & -3724.x42192 & -11.37x060 &  -6.2x453  &  -0.02x903 &   0.06x215 & -3742.x005(16) \\
 &  8 & -4445.x63274 & -12.81x245 &  -9.1x74   &   0.00x327 &   0.06x207 & -4467.x554(31) \\
 &  9 & -5041.x00923 & -13.99x389 & -12.7x97   &   0.04x705 &   0.05x647 & -5067.x697(55) \\
 & 10 & -5539.x33827 & -14.97x737 & -17.1x24   &   0.10x127 &   0.04x523 & -5571.x293(91) 
  \end{tabular}
\end{ruledtabular}
\end{table*}

The third part of $\cE^{(7)}$ is given by the second order matrix elements
of the form
\cite{pachucki:06:prl:he}
\begin{align} 
\cE^{(7)}_{\rm sec} =  &\ 
  2\lbr H^{(4)}_{\rm fs}\frac{1}{(E_0-H_0)'}H^{(5)}_{\rm nlog}\rbr 
\nonumber \\ &
  + 2\lbr H^{(4)}\frac{1}{(E_0-H_0)'}H^{(5)}_{\rm fs}\rbr \,,
\end{align} 
where $H^{(4)} = H^{(4)}_{\rm fs}+H^{(4)}_{\rm nfs}$ is the Breit-Pauli Hamiltonian
(see Eq.~(6) of Ref.~\cite{pachucki:09:hefs}),
$H^{(5)}_{\rm fs}$ is the amm correction to $H^{(4)}_{\rm fs}$
and
\begin{align} 
 H^{(5)}_{\rm nlog} = - \frac{7}{6\,\pi\,r^3}
+ \frac{38 Z}{45}\, \left[\delta^3(r_1)+\delta^3(r_2)\right] \,.
\end{align}  

The fourth part of $\cE^{(7)}$ is the
low-energy contribution $\cE_{L}^{(7)}$ that can be interpreted as 
the relativistic correction to the Bethe logarithm. It is given by
\cite{pachucki:00:jpb}
\begin{align}
&\ \cE_{L}^{(7)} =
 \nonumber \\ & 
 -\frac{2}{3\,\pi}\,
\delta\, \Bigl< (\bfp_1+\bfp_2)\,(H_0-E_0)
\ln\left[\frac{2(H_0-E_0)}{Z^2}\right]
(\bfp_1+\bfp_2)\Bigr>
 \nonumber \\ & 
+\frac{i\,Z^2}{3\,\pi}
\Bigl< \left(\frac{\bfr_1}{r_1^3}+\frac{\bfr_2}{r_2^3}\right) \times 
     \frac{\bsigma_1+\bsigma_2}{2}
\ln\left[\frac{2(H_0-E_0)}{Z^2}\right]
\left(\frac{\bfr_1}{r_1^3}+\frac{\bfr_2}{r_2^3}\right) \Bigr>\,, 
\label{bethe}
\end{align}
where $\delta \lbr \ldots \rbr$ denotes the first-order perturbation of
the matrix element $\langle\ldots\rangle$ by $H^{(4)}_{\rm fs}$.

Our calculational results for the corrections of order $m\alpha^7$ and 
$m^2\alpha^6/M$ are listed in Table~\ref{tab:E7}. For the 
logarithmic part $\cE^{(7)}_{\rm log}$, our results fully confirm 
the previous calculation \cite{zhang:96:prl}. 
The recoil correction $\cE^{(6)}_M$ and a part of 
the second-order contribution $\cE^{(7)}_{\rm sec}$ were calculated for
helium by Drake \cite{drake:02:cjp}. Our results agree with those of Drake for
the second-order part but differ by about 5\% for the recoil correction. The difference
entails a small shift of about 0.5~kHz for the $\nu_{01}$ and
$\nu_{12}$ intervals. 
The helium results listed in Table~\ref{tab:E7} differ from those reported
by us previously \cite{pachucki:09:hefs} only in the 
Bethe logarithm part $\cE^{(7)}_L$. By checking the hydrogenic
limit for this correction, we found that our previous evaluation
\cite{pachucki:09:hefs} contained a mistake. Its source was a 
term missing in the final expressions for $E_{L1}$. 
More specifically, $\ln K$ and $\ln \kappa$ in Eqs.~(168) and (173) of that work 
should be replaced by $\ln (2K/Z^2)$ and $\ln (2\kappa/Z^2)$, respectively. (To
note, the term in question was correctly accounted for in the original calculation 
\cite{pachucki:00:jpb}.) This term increases the theoretical values of the $\nu_{01}$ and
$\nu_{12}$ intervals by 6.1 and 1.6~kHz, respectively.

Table~\ref{tab:E7} also presents the results for the high-$Z$ limit of
individual $m\alpha^7$ corrections. This limit was evaluated numerically by fitting the
$1/Z$ expansion of our numerical data and compared to the analytical results
known from the hydrogen theory \cite{jentschura:96}.
A remarkable feature of the $m\alpha^7$ corrections is
their strong $Z$ dependence. Table~\ref{tab:E7} demonstrates that for 
the largest $Z$ studied, 
the values of $\cE^{(7)}_{\rm log}$ and $\cE^{(7)}_{L}$ are still very different
from their hydrogenic limits (even the sign is often opposite). 

Combining the results presented in Table~\ref{tab:E7} with 
the contributions of lower orders from our previous investigation 
\cite{yerokhin:10:helike}, we obtain total theoretical values of the
fine-structure intervals in light helium-like atoms summarized in
Table~\ref{tab:total}. The uncertainties quoted in the table
are due to uncalculated effects to order $m\alpha^8$. These
effects were estimated by scaling the
$m\alpha^6$ correction by the factor of $(\Za)^2$. For helium, the estimates
for the $\nu_{01}$ and $\nu_{12}$ intervals were obtained by taking the $m\alpha^6$
correction for $\nu_{02}$. In all other instances, 
the $m\alpha^6$ correction for the corresponding interval was taken.
It is remarkable that in all the cases except helium, the theoretical accuracy is 
significantly (usually by a factor of $1/Z$) better for the $\nu_{02}$ 
interval than for $\nu_{01}$ and $\nu_{12}$. This is due to the absence of the
leading term in the $1/Z$ expansion of the $m\alpha^6$ 
correction (and some others) for the $\nu_{02}$ interval.

We note that the present calculation is performed for a spinless nucleus.
For a nucleus with spin, the hyperfine splitting (hfs) 
can usually be evaluated separately and employed for an
experimental determination of the fine structure. This procedure, however, 
ignores the mixing between the hfs and the fine structure. So, more accurate
calculations should account for both effects simultaneously.

The comparison with experiment is summarized in Table~\ref{tab:compar}. The
agreement between theory and experiment is usually very good. The only
significant discrepancy is for Be$^{2+}$, where the difference 
amounts to $1.7$ standard deviations ($\sigma$) 
for $\nu_{12}$ and $3.5\,\sigma$ for $\nu_{02}$. Our
result for the $\nu_{01}$ interval of helium agrees well with the experimental 
values \cite{zelevinsky:05,giusfredi:05,george:01}. For the $\nu_{12}$ interval, our
theory is by about $2\sigma$ away from the values obtained in 
Refs.~\cite{zelevinsky:05,castillega:00} but
in agreement with the latest measurement by Hessels and coworkers~\cite{borbely:09}. 

Assuming the validity of the theory, we
combine the theoretical prediction 
for the $\nu_{01}$ interval in helium with the
experimental result \cite{zelevinsky:05} and obtain
the following value of the fine structure constant,
\begin{equation} \label{alpha}
\alpha^{-1}({\rm He}) = 137.036\,001\,1\,(39)_{\rm theo}\,(16)_{\rm exp}\,,
\end{equation}
which is accurate to 31~ppb and agrees with the more precise results of 
Refs.~\cite{hanneke:08,cadoret:08}. The theoretical uncertainty of the above
value of $\alpha$ is more than twice larger than the experimental one. 
In order to improve the theoretical accuracy, one has to calculate
the $m\alpha^8$ correction. Its complete evaluation is 
extremely difficult. One can hope, however, to identify the dominant
part of this effect, since most of $m\alpha^8$ operators should be negligible. 
This task is simpler to accomplish for the $\nu_{02}$ interval, since 
the effects of the triplet-singlet mixing are absent in this case. 
It is also possible to estimate the $m\alpha^8$
correction from an independent measurement 
for a different $Z$. So, an accurate
experimental determination of the $\nu_{02}$ interval in a light helium-like
ion (preferably, $^{12}$C$^{4+}$ since it has a spinless nucleus) would yield an estimate for 
the $m\alpha^8$ term in helium with a 50\% accuracy,
thus reducing the theoretical uncertainty of this interval
by a factor of 2.

\begin{table}
\caption{Comparison of theoretical and experimental results for
the fine-structure intervals of helium-like atoms.
Units are MHz for He and Li$^+$ and cm$^{-1}$ for other atoms.
\label{tab:compar}} 
\begin{ruledtabular}
  \begin{tabular}{cl..r}
 $(J,J')$ &  
 $Z$ &  
\multicolumn{1}{c}{Present work}        
             &  \multicolumn{1}{c}{Experiment}        
                         &  Ref.        \\
    \hline\\[-5pt]
(0,1)
&  2 &   29\,616.x952\,3(17)  &   29\,616.x951\,66(70) & \cite{zelevinsky:05}\\
&    &                        &   29\,616.x952\,7(10)  & \cite{giusfredi:05}\\
&    &                        &   29\,616.x950\,9(9)   & \cite{george:01} \\
&  3 &  155\,704.x584(48)     &  155\,704.x27(66)      & \cite{riis:94} \\
&  4 &   11.5x57\,756(33)     &    11.5x58\,6(5)       & \cite{scholl:93} \\
&  5 &   16.1x98\,21(29)      &    16.2x03(18)         & \cite{dinneen:91} \\
&  7 &   -8.6x73\,1(67)       &    -8.6x70\,7(7)       & \cite{thompson:98} 
\\[5pt]
(1,2)
&  2 &    2291x.178\,9(17)  &    2\,291.x177\,53(35) & \cite{borbely:09} \\
&    &                      &    2\,291.x175\,59(51) & \cite{zelevinsky:05}\\
&    &                      &    2\,291.x175\,9(10)  & \cite{castillega:00}\\
&  9 &  -957.8x86(79)       &  -957.8x73\,0(12)      & \cite{myers:99:F} 
\\[5pt]
(0,2)
&  3 &    93\,025.x266(34)   &  93\,025.x86(61)       & \cite{riis:94} \\
&  4 &     -3.3x34\,663(10)  &    -3.3x36\,4(5)       & \cite{scholl:93} \\
&  5 &    -36.4x63\,787(66)  &   -36.4x57(16)         & \cite{dinneen:91} 
  \end{tabular}
\end{ruledtabular}
\end{table}

In summary, our present study concludes the evaluation of the $m\alpha^7$
correction to the fine structure of light helium-like atoms and
resolves the discrepancy between theory and experiment reported  in the
literature. The theoretical values agree with the
latest experimental results for helium, as well as with most of the experimental 
data for helium-like ions. A combination of the theoretical and experimental results
for the $2^3P_1-2^3P_0$ interval in helium yields an independent determination of 
the fine structure constant $\alpha$ accurate to 31~ppb. The
precision will be increased further when more accurate estimates of the
higher-order effects are obtained from theoretical or experimental studies.

Support by NIST through Precision Measurement Grant PMG 60NANB7D6153
is gratefully acknowledged. V.A.Y. was also supported by RFBR (grant No.~10-02-00150-a).

\end{document}